\documentclass[aps,prl,twocolumn,superscriptaddress,showpacs,preprintnumbers,amsmath,amssymb]{revtex4-1}
\RequirePackage{lineno}

\usepackage{lipsum}
\usepackage{hyperref}
\usepackage{graphicx} 
\usepackage{dcolumn}  
\usepackage{graphicx} 
\usepackage{dcolumn}  
\usepackage{xspace}
\usepackage{cleveref}
\usepackage{amsmath} 
\usepackage{subfigure}
\usepackage{booktabs}
\usepackage{longtable}
\usepackage{dcolumn}

\usepackage[usenames,dvipsnames,svgnames,table]{xcolor}

\hypersetup{colorlinks=true, citecolor=Blue, urlcolor=Blue, linkcolor=Blue}

\graphicspath{{ps}}

\newcommand*\patchAmsMathEnvironmentForLineno[1]{%
  \expandafter\let\csname old#1\expandafter\endcsname\csname #1\endcsname
  \expandafter\let\csname oldend#1\expandafter\endcsname\csname end#1\endcsname
  \renewenvironment{#1}%
     {\linenomath\csname old#1\endcsname}%
     {\csname oldend#1\endcsname\endlinenomath}}%
\newcommand*\patchBothAmsMathEnvironmentsForLineno[1]{%
  \patchAmsMathEnvironmentForLineno{#1}%
  \patchAmsMathEnvironmentForLineno{#1*}}%
\AtBeginDocument{%
\patchBothAmsMathEnvironmentsForLineno{equation}%
\patchBothAmsMathEnvironmentsForLineno{align}%
\patchBothAmsMathEnvironmentsForLineno{flalign}%
\patchBothAmsMathEnvironmentsForLineno{alignat}%
\patchBothAmsMathEnvironmentsForLineno{gather}%
\patchBothAmsMathEnvironmentsForLineno{multline}%
}

 \makeatletter
 \DeclareRobustCommand*{\bfseries}{%
   \not@math@alphabet\bfseries\mathbf
   \fontseries\bfdefault\selectfont
   \boldmath
 }

\begin{document}
\vspace*{0.5cm}

%
\preprint{\vbox{ \hbox{   }
		\hbox{Belle Preprint 2016-15}
		\hbox{KEK Preprint 2016-54}
}}

\newcommand{\phit}{\textless phi-t\textgreater\textsuperscript\textregistered\,}
\newcommand{\neurobayes}{NeuroBayes\textsuperscript\textregistered\,}

%
\newcommand{\mevc}{\ensuremath{\mathrm{MeV}/c^2}\xspace}
\newcommand{\gevc}{\ensuremath{\mathrm{GeV}/c^2}\xspace}
\newcommand{\gev}{\ensuremath{\mathrm{GeV}}\xspace}
\newcommand{\yvs}{$\Upsilon(5S)$ }
\newcommand{\sm}{Standard Model\xspace}
\newcommand{\SM}{Standard Model\xspace}
\newcommand{\NP}{new physics\xspace}
\newcommand{\B}{$B$ meson\xspace}
\newcommand{\Bs}{$B$ mesons\xspace}
\newcommand{\Bjpsi}{$B\to K^\ast J/\psi$\xspace}
\newcommand{\Bpsi}{$B\to K^\ast \psi(2S)$\xspace}
\newcommand{\Bto}[1]{
	\ifnum #1 = 521343 \ensuremath{B^+ \to K^+ e^+ e^- \xspace}\fi
	\ifnum #1 = 521347 \ensuremath{B^+ \to K^+ \mu^+ \mu^- \xspace}\fi
	\ifnum #1 = 521345 \ensuremath{B^+ \to K^{\ast +} e^+ e^- \xspace}\fi
	\ifnum #1 = 521349 \ensuremath{B^+ \to K^{\ast +} \mu^+ \mu^- \xspace}\fi
	\ifnum #1 = 511332 \ensuremath{B^0 \to K^0 e^+ e^- \xspace}\fi
	\ifnum #1 = 511336 \ensuremath{B^0 \to K^0 \mu^+ \mu^- \xspace}\fi
	\ifnum #1 = 511335 \ensuremath{B^0 \to K^{\ast 0} e^+ e^- \xspace}\fi
	\ifnum #1 = 511339 \ensuremath{B^0 \to K^{\ast 0}  \mu^+ \mu^- \xspace}\fi
}
\newcommand{\Kto}[1]{
	\ifnum #1 = 313421 \ensuremath{K^{\ast 0} \to K_S \pi^0}\fi
	\ifnum #1 = 313532 \ensuremath{K^{\ast 0} \to K^+ \pi^-}\fi
	\ifnum #1 = 323432 \ensuremath{K^{\ast +} \to K^+ \pi^0}\fi
	\ifnum #1 = 323521 \ensuremath{K^{\ast +} \to K_S \pi^+}\fi
}
\newcommand{\mbc}{\ensuremath{M_\mathrm{bc}}\xspace}
\newcommand{\de}{\ensuremath{\Delta E}\xspace}
\newcommand{\ctl}{$\cos\theta_\ell$\xspace}
\newcommand{\ctk}{$\cos\theta_K$\xspace}
\newcommand{\ph}{$\phi$\xspace}
\newcommand{\nbout}{$NB_{out}$}
\newcommand*{\factor}{0.95}
\newcommand*{\factorh}{0.48}
\newcommand*{\factort}{0.315}
\newcommand*{\factorq}{0.24}
\newcommand{\yfs}{$\Upsilon(4S)$\xspace}
\newcommand{\bkll}{$B \to K^{(\ast)} \ell^+ \ell^-$\xspace}
\newcommand{\bkstll}{$B \to K^{\ast} \ell^+ \ell^-$\xspace}
\newcommand{\bkllzero}{\ensuremath{B^0 \to K^\ast(892)^0  \ell^+ \ell^-\xspace}}
\newcommand{\bsll}{$b \to s \ell^+ \ell^-$\xspace}
\newcommand{\bktt}{\ensuremath{B^+ \to K^+ \tau^+ \tau^-}\xspace}
\newcommand{\bkee}{\ensuremath{B^+ \to K^+ e^+ e^-}\xspace}
\newcommand{\bkttp}{\ensuremath{B^+ \to K^+ \tau^+ \tau^-}\xspace}
\newcommand{\bkllp}{$B \to K^{(\ast)} \ell^+ \ell^-$\xspace}
\newcommand{\bmn}{$B$ meson\xspace}
\newcommand{\bms}{$B$ mesons\xspace}
\newcommand{\kast}{\ensuremath{K^\ast}\xspace}
\newcommand{\bc}{$B$ candidate\xspace}
\newcommand{\cms}{center-of-mass\xspace}
\newcommand{\nbb}{772 million\xspace}
\newcommand{\eecl}{\ensuremath{E_{ECL}}\xspace}
\newcommand{\btag}{$B_{tag}$\xspace}
\newcommand{\upperlimm}{${\cal B}(B^+ \to K^+ \tau^+ \tau^-)< 3.17\times 10^{-4}$\xspace}
\newcommand{\cutbdt}{$c_{BDT}=0.553$\xspace}
\newcommand{\cuteecl}{$c_{E}=0.206~\mathrm{GeV}$\xspace}
\newcommand{\nbkg}{$8.83$\xspace}
\newcommand{\effsig}{$\epsilon_s=2.709\times10^{-5}$\xspace}
\newcommand{\upl}[1]{\ensuremath{{\cal B}(B^+ \to K^+ \tau^+ \tau^-)< #1}}
\newcommand{\upll}[2]{\ensuremath{{\cal B}(B^+ \to K^+ \tau^+ \tau^-)< #1 \times 10^{- #2}}}
\newcommand{\uplll}[3]{\ensuremath{{\cal B}(#1)< #2 \times 10^{- #3}}}
\newcommand{\upllll}[4]{\ensuremath{{\cal B}(#1)^{#2}< #3 \times 10^{- #4}}}
\newcommand{\br}[4]{\ensuremath{{\cal B}(#1)^{#2}= #3 \times 10^{- #4}}}
\newcommand*\diff{\mathop{}\!\mathrm{d}}
\newcommand*\Diff[1]{\mathop{}\!\mathrm{d^#1}}
\newcommand{\gevsq}{\ensuremath{~\mathrm{GeV}^2/c^4}\xspace}
\newcommand{\fitcomponents}{Combinatorial (shaded
	 blue), signal (red filled) and total (solid) fit distributions are superimposed on the data points.}
\newcommand{\fitprojectioncaption}[2]{Projections for the fit result of $P_{#1}'$ in bin #2.  Fit to the \mbc sideband for the determination of the background shape (top) and signal region (bottom) are displayed. \fitcomponents}
\newcommand{\ckmt}[2]{\ensuremath{V_{#1}^{\;}V_{#2}^\ast}\xspace}
\title{Lepton-Flavor-Dependent Angular Analysis of \bkstll}
\noaffiliation
\affiliation{University of the Basque Country UPV/EHU, 48080 Bilbao}
\affiliation{Beihang University, Beijing 100191}
\affiliation{University of Bonn, 53115 Bonn}
\affiliation{Budker Institute of Nuclear Physics SB RAS, Novosibirsk 630090}
\affiliation{Faculty of Mathematics and Physics, Charles University, 121 16 Prague}
\affiliation{Chonnam National University, Kwangju 660-701}
\affiliation{University of Cincinnati, Cincinnati, Ohio 45221}
\affiliation{Deutsches Elektronen--Synchrotron, 22607 Hamburg}
\affiliation{University of Florida, Gainesville, Florida 32611}
\affiliation{Justus-Liebig-Universit\"at Gie\ss{}en, 35392 Gie\ss{}en}
\affiliation{SOKENDAI (The Graduate University for Advanced Studies), Hayama 240-0193}
\affiliation{Hanyang University, Seoul 133-791}
\affiliation{University of Hawaii, Honolulu, Hawaii 96822}
\affiliation{High Energy Accelerator Research Organization (KEK), Tsukuba 305-0801}
\affiliation{J-PARC Branch, KEK Theory Center, High Energy Accelerator Research Organization (KEK), Tsukuba 305-0801}
\affiliation{IKERBASQUE, Basque Foundation for Science, 48013 Bilbao}
\affiliation{Indian Institute of Science Education and Research Mohali, SAS Nagar, 140306}
\affiliation{Indian Institute of Technology Bhubaneswar, Satya Nagar 751007}
\affiliation{Indian Institute of Technology Guwahati, Assam 781039}
\affiliation{Indian Institute of Technology Madras, Chennai 600036}
\affiliation{Indiana University, Bloomington, Indiana 47408}
\affiliation{Institute of High Energy Physics, Chinese Academy of Sciences, Beijing 100049}
\affiliation{Institute of High Energy Physics, Vienna 1050}
\affiliation{Institute of Mathematical Sciences, Chennai 600113}
\affiliation{INFN - Sezione di Torino, 10125 Torino}
\affiliation{J. Stefan Institute, 1000 Ljubljana}
\affiliation{Kanagawa University, Yokohama 221-8686}
\affiliation{Institut f\"ur Experimentelle Kernphysik, Karlsruher Institut f\"ur Technologie, 76131 Karlsruhe}
\affiliation{Kennesaw State University, Kennesaw, Georgia 30144}
\affiliation{Department of Physics, Faculty of Science, King Abdulaziz University, Jeddah 21589}
\affiliation{Korea Institute of Science and Technology Information, Daejeon 305-806}
\affiliation{Korea University, Seoul 136-713}
\affiliation{Kyungpook National University, Daegu 702-701}
\affiliation{\'Ecole Polytechnique F\'ed\'erale de Lausanne (EPFL), Lausanne 1015}
\affiliation{P.N. Lebedev Physical Institute of the Russian Academy of Sciences, Moscow 119991}
\affiliation{Faculty of Mathematics and Physics, University of Ljubljana, 1000 Ljubljana}
\affiliation{Ludwig Maximilians University, 80539 Munich}
\affiliation{University of Maribor, 2000 Maribor}
\affiliation{Max-Planck-Institut f\"ur Physik, 80805 M\"unchen}
\affiliation{School of Physics, University of Melbourne, Victoria 3010}
\affiliation{University of Miyazaki, Miyazaki 889-2192}
\affiliation{Moscow Physical Engineering Institute, Moscow 115409}
\affiliation{Moscow Institute of Physics and Technology, Moscow Region 141700}
\affiliation{Graduate School of Science, Nagoya University, Nagoya 464-8602}
\affiliation{Kobayashi-Maskawa Institute, Nagoya University, Nagoya 464-8602}
\affiliation{Nara Women's University, Nara 630-8506}
\affiliation{National Central University, Chung-li 32054}
\affiliation{National United University, Miao Li 36003}
\affiliation{Department of Physics, National Taiwan University, Taipei 10617}
\affiliation{H. Niewodniczanski Institute of Nuclear Physics, Krakow 31-342}
\affiliation{Nippon Dental University, Niigata 951-8580}
\affiliation{Niigata University, Niigata 950-2181}
\affiliation{Novosibirsk State University, Novosibirsk 630090}
\affiliation{Osaka City University, Osaka 558-8585}
\affiliation{Pacific Northwest National Laboratory, Richland, Washington 99352}
\affiliation{University of Pittsburgh, Pittsburgh, Pennsylvania 15260}
\affiliation{Theoretical Research Division, Nishina Center, RIKEN, Saitama 351-0198}
\affiliation{University of Science and Technology of China, Hefei 230026}
\affiliation{Showa Pharmaceutical University, Tokyo 194-8543}
\affiliation{Soongsil University, Seoul 156-743}
\affiliation{Stefan Meyer Institute for Subatomic Physics, Vienna 1090}
\affiliation{Sungkyunkwan University, Suwon 440-746}
\affiliation{School of Physics, University of Sydney, New South Wales 2006}
\affiliation{Department of Physics, Faculty of Science, University of Tabuk, Tabuk 71451}
\affiliation{Tata Institute of Fundamental Research, Mumbai 400005}
\affiliation{Excellence Cluster Universe, Technische Universit\"at M\"unchen, 85748 Garching}
\affiliation{Department of Physics, Technische Universit\"at M\"unchen, 85748 Garching}
\affiliation{Toho University, Funabashi 274-8510}
\affiliation{Department of Physics, Tohoku University, Sendai 980-8578}
\affiliation{Earthquake Research Institute, University of Tokyo, Tokyo 113-0032}
\affiliation{Department of Physics, University of Tokyo, Tokyo 113-0033}
\affiliation{Tokyo Institute of Technology, Tokyo 152-8550}
\affiliation{Tokyo Metropolitan University, Tokyo 192-0397}
\affiliation{University of Torino, 10124 Torino}
\affiliation{Virginia Polytechnic Institute and State University, Blacksburg, Virginia 24061}
\affiliation{Wayne State University, Detroit, Michigan 48202}
\affiliation{Yamagata University, Yamagata 990-8560}
\affiliation{Yonsei University, Seoul 120-749}
%
  \author{S.~Wehle}\affiliation{Deutsches Elektronen--Synchrotron, 22607 Hamburg} 
  \author{C.~Niebuhr}\affiliation{Deutsches Elektronen--Synchrotron, 22607 Hamburg} 
   \author{S.~Yashchenko}\affiliation{Deutsches Elektronen--Synchrotron, 22607 Hamburg} 
  \author{I.~Adachi}\affiliation{High Energy Accelerator Research Organization (KEK), Tsukuba 305-0801}\affiliation{SOKENDAI (The Graduate University for Advanced Studies), Hayama 240-0193} 
  \author{H.~Aihara}\affiliation{Department of Physics, University of Tokyo, Tokyo 113-0033} 
  \author{S.~Al~Said}\affiliation{Department of Physics, Faculty of Science, University of Tabuk, Tabuk 71451}\affiliation{Department of Physics, Faculty of Science, King Abdulaziz University, Jeddah 21589} 
  \author{D.~M.~Asner}\affiliation{Pacific Northwest National Laboratory, Richland, Washington 99352} 
  \author{V.~Aulchenko}\affiliation{Budker Institute of Nuclear Physics SB RAS, Novosibirsk 630090}\affiliation{Novosibirsk State University, Novosibirsk 630090} 
  \author{T.~Aushev}\affiliation{Moscow Institute of Physics and Technology, Moscow Region 141700} 
  \author{R.~Ayad}\affiliation{Department of Physics, Faculty of Science, University of Tabuk, Tabuk 71451} 
% ...
\collaboration{The Belle Collaboration}

\noaffiliation

\begin{abstract}
We present a measurement of  angular observables and a test of lepton flavor universality   in the \bkstll  decay, where $\ell$ is either $e$ or $\mu$.
The analysis is performed on a data sample corresponding to an integrated luminosity of $711~\mathrm{fb}^{-1}$  containing $772\times 10^{6}$ $B\bar B$ pairs, collected at the \yfs resonance with the Belle detector at the asymmetric-energy $e^+e^-$ collider KEKB.
The result is consistent with  Standard Model (SM) expectations,
where the largest discrepancy from a SM prediction is observed in the muon modes with a local significance of $2.6\sigma$.
\end{abstract}


\maketitle


\clearpage


In this Letter,  a measurement of angular observables and a test of lepton flavor universality (LFU)  in the \bkstll decay  is presented, where $\ell=e,\mu$.
The  \bkstll  decay 
involves the quark transition $b\to s \ell^+ \ell^-$, a flavor-changing neutral current  that is forbidden at tree level in the Standard Model (SM).
Various extensions to the SM predict contributions from new physics (NP), which can interfere with the SM amplitudes. 
In recent years, several measurements have shown deviations from the SM  in this particular decay \cite{lhcb2,Aaij:2015esa,rk}. 
Global analyses of $B$ decays hint at lepton-flavor non-universality,
in which case muon modes would have larger contributions from NP
than electron modes \cite{globalbsll,globalstraub}.
%
%
%

The decay can be   described kinematically by three angles $\theta_\ell$, $\theta_K$, $\phi$ and the invariant mass squared of the lepton pair $q^2\equiv M^2_{\ell\ell}c^2$. 
The angle $\theta_\ell$ is defined as the angle between the direction of $\ell^+ ~(\ell^-)$ and the direction opposite the $B~(\bar B)$ in the dilepton rest frame.
The angle $\theta_K$ is defined as the angle between the direction of the kaon and the  direction opposite the $B~(\bar B)$ in the $K^\ast$ rest frame.
Finally, the angle $\phi$ is defined as the angle between the  plane formed by the $\ell^+\ell^-$ system  and the $K^\ast$ decay plane in the $B ~(\bar B)$ rest frame.
The differential decay rate can be parametrized using  definitions presented in Ref.~\cite{Altmannshofer:2008dz}
 by
\begin{widetext}
\begin{align}
\frac{1}{\textrm{d}\Gamma/\textrm{d} q^2}
\frac{\textrm{d}^4\Gamma}{\textrm{d}\cos\theta_\ell\; \textrm{d}\cos\theta_K\; \textrm{d}\phi\; \textrm{d}q^2}
= 
&\frac{9}{32\pi}
\left[
\frac{3}{4} (1 - F_L) \sin^2\theta_K 
+F_L\cos^2\theta_K \right.
+\frac{1}{4}(1-F_L)\sin^2\theta_K\cos2\theta_\ell \nonumber \\
&-F_L \cos^2\theta_K\cos 2\theta_\ell + S_3 \sin^2\theta_K \sin^2\theta_\ell\cos 2\phi 
 + S_4 \sin 2\theta_K \sin 2\theta_\ell \cos\phi  \nonumber  \\
&+ S_5 \sin 2\theta_K\sin\theta_\ell\cos\phi 
+S_6 \sin^2\theta_K\cos\theta_\ell + S_7 \sin 2\theta_K \sin\theta_\ell\sin\phi \nonumber \\ 
&+  S_8 \sin 2\theta_K\sin 2\theta_\ell\sin\phi + S_9 \sin^2\theta_K\sin^2\theta_\ell\sin 2\phi \;\bigg],\label{eq:signalpdf}
\end{align}
\end{widetext}
where the  observables $F_L$ and $S_i$ are functions of $q^2$ only.
The observables $P_i'$, introduced in Ref.~\cite{DescotesGenon:2012zf} and defined as
\begin{equation}
P'_{i=4,5,6,8} = \frac{S_{j=4,5,7,8}}{\sqrt{F_L(1-F_L)}},
\end{equation}
are considered to be largely free of form-factor uncertainties \cite{p_theory}.
Any deviation from zero in the difference $Q_i = P_i^\mu - P_i^e$ would be a direct hint of   new physics \cite{DHMVeemumu}; here,  $i=4,5$ and $P{_{i}^\ell}$ refers  to  $P_{4,5}'$ in the corresponding lepton mode.
The definition of $P_i'$ values follows the LHCb convention \cite{lhcb2}.

In previous measurements of the  $P_i'$ observables  only $B^0$ decays, followed by  $K^{\ast 0}$ decays to $K^+ \pi^-$, were used \cite{lhcb2}.
This measurement also uses $B^+$ decays, where $K^{\ast+}\to  K^+\pi^0$  or $K^0_S\pi^+$.  
In total, the decay modes \Bto{511339},  \Bto{521349}, \Bto{511335}, and   \Bto{521345} are reconstructed, where the inclusion of charge-conjugate states is implied if not explicitly stated.
The full $\Upsilon(4S)$ data sample is used containing $772\times 10^{6}$ $B\bar B$ pairs recorded with the Belle detector \cite{BelleDetektor} at the asymmetric-energy $e^+e^-$ collider KEKB \cite{kekb}.
The Belle detector is a large-solid-angle magnetic
spectrometer that consists of a silicon vertex detector,
a 50-layer central drift chamber (CDC), an array of
aerogel threshold Cherenkov counters (ACC),  
a barrel-like arrangement of time-of-flight
scintillation counters (TOF), and an electromagnetic calorimeter
comprised of CsI(Tl) crystals (ECL) located inside 
a superconducting solenoid coil that provides a 1.5~T
magnetic field.  An iron flux-return located outside of
the coil is instrumented to detect $K_L^0$ mesons and to identify
muons (KLM).  The detector
is described in detail elsewhere~\cite{BelleDetektor}.
This analysis is validated and optimized using simulated  Monte Carlo (MC) data samples. 
EvtGen \cite{evtgen} and PYTHIA \cite{pythia} are used to simulate the particle decays.
Final-state radiation is calculated by the PHOTOS package \cite{photos}.
The  detector response is simulated with  GEANT3 \cite{geant}.

For all charged tracks, impact parameter requirements  are applied with respect to the nominal interaction point along the beam direction  ($|dz| <5.0~\textrm{cm}$)  and in the transverse plane   ($dr <1.0~\textrm{cm}$).
For electrons, muons, $K^+$, and $\pi^+$, a particle identification likelihood is calculated from the energy loss  in the CDC ($\mathrm{d}E/\mathrm{d}x$),  time-of-flight  measurements in the TOF, the response of the ACC, the transverse  shape and size of the showers in the ECL and information about hits in the KLM.
For electrons, energy loss from bremsstrahlung is recovered by adding to the candidate the momenta of photons in a cone of $0.05$ radians around the initial direction of the charged track.
$K_S^0$ candidates  are reconstructed from pairs of  oppositely-charged tracks (treated as pions) and selected based on vertex fit quality.
$\pi^0$ mesons are reconstructed from photon pairs  with the requirement $E_\gamma>30~\mathrm{MeV}$ and $115~\mevc<M_{\gamma\gamma} < 153~\mathrm{MeV}/c^2$.
$K^\ast$ candidates are formed from $K^+\pi^-$,  $K^+\pi^0$ and  $K^0_S\pi^+$ combinations that satisfy the requirements on   invariant mass  of $0.6~\textrm{GeV}/c^2 < M_{K\pi} < 1.4~\textrm{GeV}/c^2$  and on  vertex fit quality (to suppress background).
The $K^{\ast}$ candidates are  combined  with oppositely charged lepton pairs to form \bmn candidates, where the charge of the kaon or pion defines the charge or flavor of the \bmn. 
The particle selection criteria lead to combinatorial background that is  suppressed by applying requirements on the  beam-energy constrained  mass, $ M_\textrm{bc} =  \sqrt{E^2_{\mathrm{beam}}/c^4 - |\vec p_B|^2/c^2}$, and the energy difference, $\Delta E =  E_B - E_{\mathrm{beam}}$, where $E_B$ and $\vec p_B$ are the energy and  momentum, respectively, of the reconstructed candidate in the $\Upsilon(4S)$ rest frame and $E_{\mathrm{beam}}$ is the beam energy in the center-of-mass frame.
Correctly reconstructed candidates are centered at the nominal $B$ mass in \mbc  and at zero in $\Delta E$.
Candidates that satisfy $ 5.22~\gevc < M_\mathrm{bc}  <~5.30~\mathrm{GeV}/c^2 $ and $ -0.10 \ (-0.05)~\gev <\Delta E < ~0.05~\mathrm{GeV}$ for the electron (muon) modes  are retained.
Large irreducible background contributions  arise from charmonium decays $B\to J/\psi K^{\ast} $ and $B\to \psi(2S) K^{\ast} $, in which the $c\bar c$ state decays into two leptons. 
These decays are vetoed with the requirements $-0.25~ (-0.15)~\mathrm{GeV}/c^2 <  M_{\ell\ell} - m_{J/\psi}< 0.08~\mathrm{GeV}/c^2$ and $-0.20 ~ (-0.10)~\mathrm{GeV}/c^2 <  M_{\ell\ell} - m_{\psi(2S)}< 0.08~\mathrm{GeV}/c^2$ for the electron (muon) modes. 
In the electron case, the veto is applied twice: with and without the bremsstrahlung-recovery treatment. 
Di-electron background from photon conversions ($\gamma \to e^+ e^-$) and $\pi^0$ Dalitz decays ($\pi^0\to e^+e^-\gamma$) is rejected by  requiring $M_{ee} > 0.14~\textrm{GeV}/c^2$.

To maximize signal efficiency and purity, neural networks are utilized sequentially from the bottom to the top of the decay chain, transferring  the output probability from each step  to the subsequent step so that the most effective selection requirements are applied in the last stage based on all information combined.
For all particle hypotheses, a neural network is trained to separate signal from background and an output value, $\mathit{o}_{\rm NB}$, is calculated for each candidate.
The classifiers for $e^\pm, \mu^\pm, K^\pm$, $K_S^{0}$, $\pi^0$, and $\pi^\pm$ are taken from the  neural-network-based full event reconstruction described in Ref.~\cite{fullrecon}.
For  $K^\ast$ selection, a classifier is trained on MC samples  using kinematic variables and vertex fit information.
The final classification is performed with a requirement on $\mathit{o}_{\rm NB}$  for each $B$ decay channel using event-shape variables (\textit{i.e.,} modified Fox-Wolfram moments \cite{ksfwm}), vertex fit information, and kinematic variables as input for the classifier.
The most important variables for the neural networks are $\Delta E$, the reconstructed mass of the $K^\ast$, the product of the network outputs of all secondary particles, and the distance between the two leptons along the beam direction $\Delta z_{\ell\ell}$.
If multiple candidates are found in an event (less than 2\% of the time), the most probable candidate is chosen based on  $\mathit{o}_{\rm NB}$.
The selection requirements for the neural networks are optimized  by maximizing the figure of merit $n_s/\sqrt{n_s + n_b}$  separately for the electron and muon channels, where $n_s$ and $n_b$ are the expected numbers of signal and background candidates, respectively,  calculated from MC.

\begin{figure}
\centering
        \subfigure{
                \includegraphics[width=0.24\textwidth]{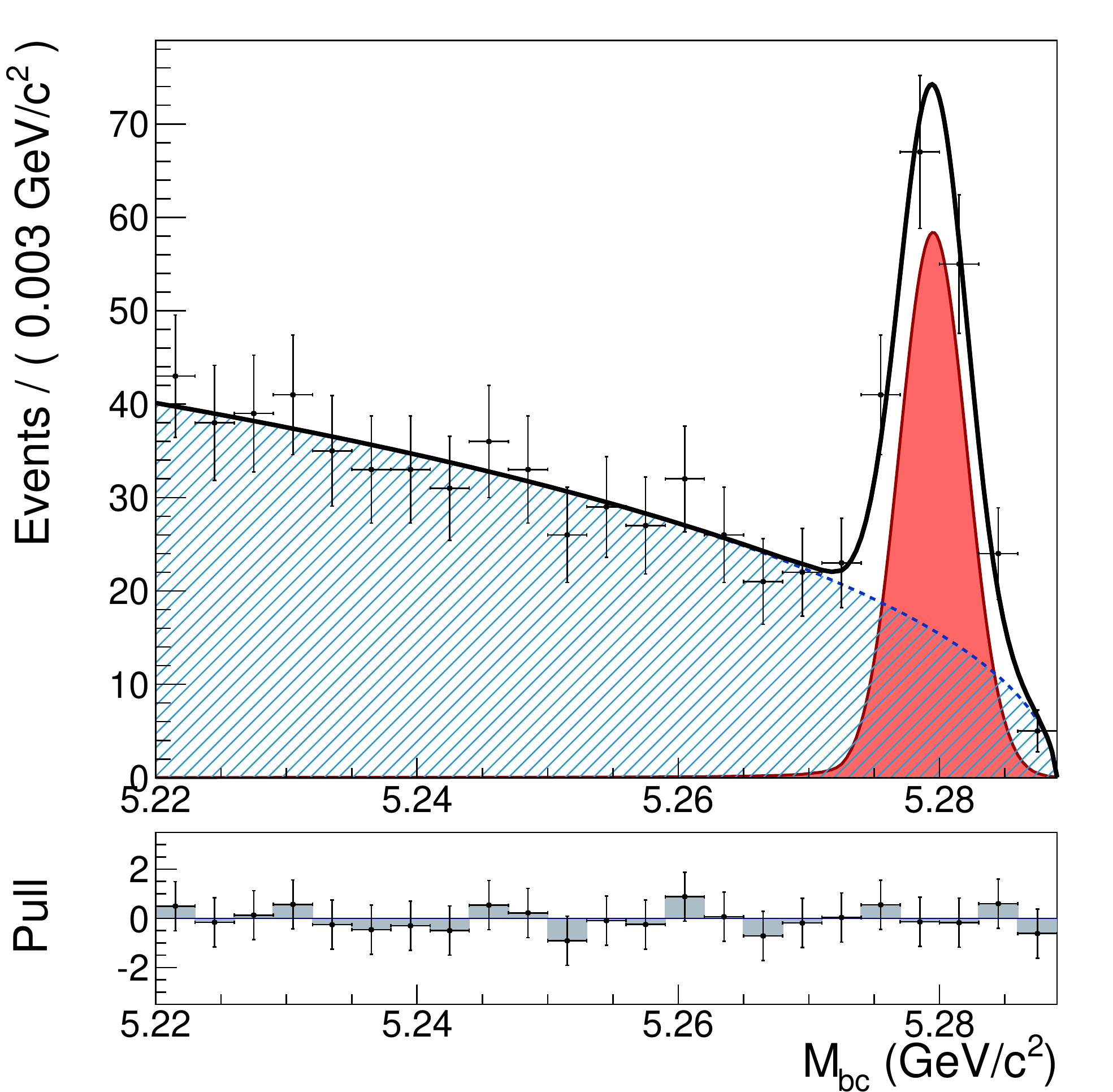}                                 
        }%
        \subfigure{
                \includegraphics[width=0.24\textwidth]{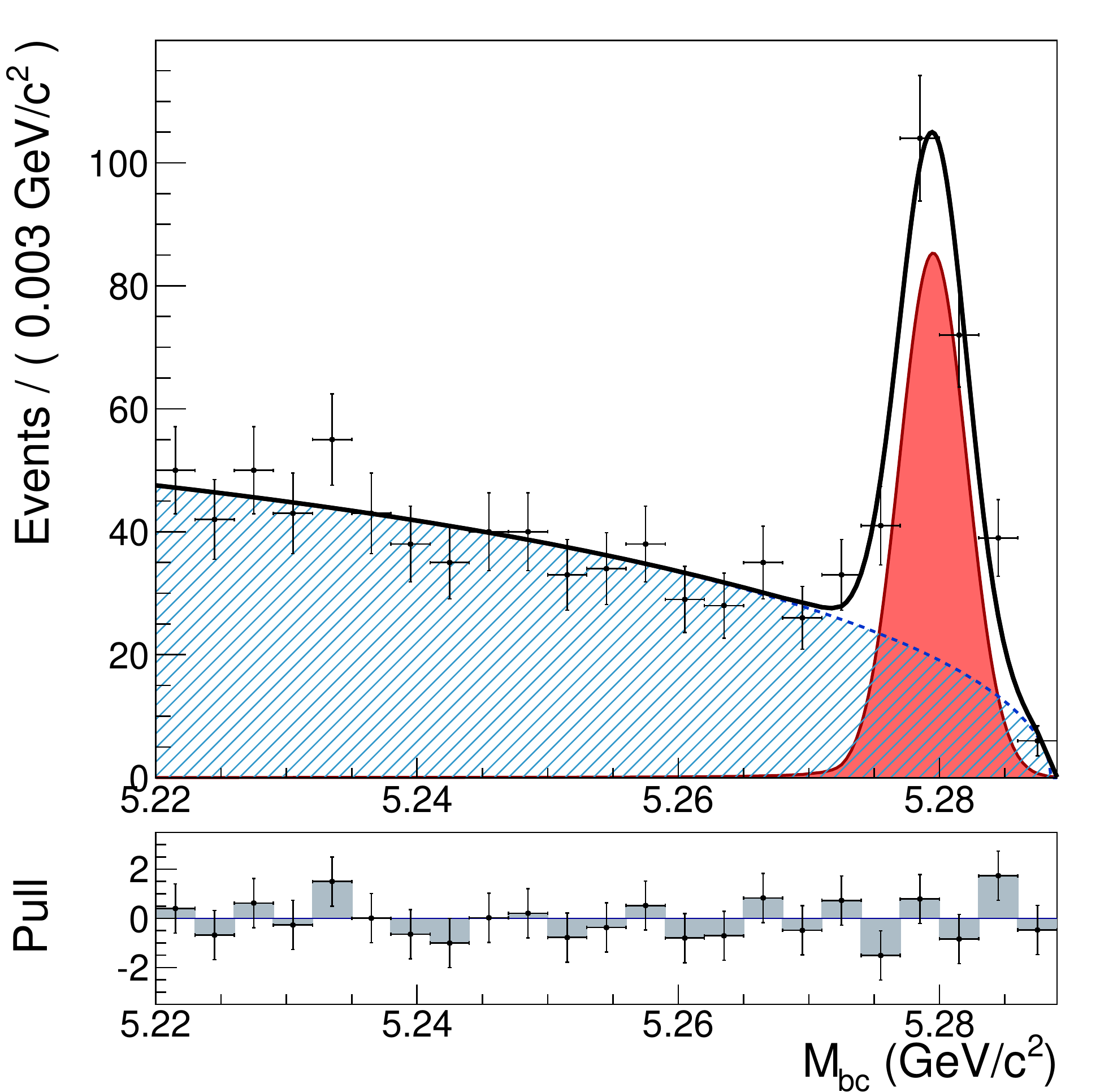}                             
        }%
    \caption{ Distribution of the beam-energy constrained mass for selected $B\to K^\ast e^+ e^-$ (left) and $B\to K^\ast \mu^+ \mu^-$ (right). Combinatorial background (shaded blue), signal (red filled) and total (solid) fit functions are superimposed on the data points}
    \label{fig:mbc_total}
\end{figure}

Signal and background yields  are extracted  by an unbinned extended maximum likelihood fit to the  \mbc  distribution of  \bkstll candidates, presented in Fig.~\ref{fig:mbc_total}, where  the signal is parametrized by a Crystal Ball function \cite{CBshape} and the background is described by an  ARGUS function \cite{ARGUSshape}.
The signal shape parameters are determined from a fit to $B\to J/\psi K^\ast$ data in the corresponding $q^2$ veto region while the background shape parameters are allowed to float in the fit.
In total $127\pm15$ and $185\pm17$ signal candidates are obtained for the electron and muon channels, respectively. 

The analysis is performed in four independent bins of $q^2$, as detailed in  \Cref{tab:result}, with an additional bin in the range	$1.0~\mathrm{GeV}^2/c^2 < q^2 < 6.0 ~\mathrm{GeV}^2/c^2$, which is favored for  theoretical predictions  \cite{Altmannshofer:2008dz}.
To make maximum use of the limited statistics, a data-transformation technique \cite{lhcb1,lhcb_phd} is applied, simplifying the differential decay rate without losing experimental sensitivity.
The transformation is applied to specific regions in the three-dimensional angular space, exploiting the symmetries of the cosine and sine functions to cancel  terms in Eq.~\ref{eq:signalpdf}.
With the following transformations to the dataset, the data are  sensitive to the observable of interest:
\begin{equation}
P_4',S_4:\;\begin{cases}
\phi \to -\phi & \text{for $\phi<0$}\\
\phi \to \pi - \phi & \text{for $\theta_\ell>\pi/2$}\\
\theta_\ell \to \pi -\theta_\ell & \text{for $\theta_\ell>\pi/2$},
\end{cases} \\
\label{eq:foldingp4} 
\end{equation}
\begin{equation}
P_5',S_5:\;\begin{cases}
\phi \to -\phi & \text{for $\phi<0$}\\
\theta_\ell \to \pi -\theta_\ell & \text{for $\theta_\ell>\pi/2$}.
\end{cases}
\label{eq:foldingp5}
\end{equation}
With this procedure, the remaining observables are the \kast longitudinal polarization, $F_L$, the transverse polarization asymmetry, $A_T^{(2)} = 2 S_3/(1-F_L)$, and  $P_4'$ or $P_5'$.
Two independent maximum likelihood fits for each bin of $q^2$ are performed to the angular distributions to extract the $P_{4,5}'$ observables.
The fits are performed using the data in the signal region of \mbc of all decay channels and separately for the electron and muon mode.
The signal (background) region is defined as $\mbc \geq 5.27~\mathrm{GeV}/c^2$ ($\mbc < 5.27~\mathrm{GeV}/c^2$).
For each measurement in $q^2$, the signal fraction is derived as a function of \mbc. 
The background angular distribution is described using the direct product of kernel density template histograms \cite{kde} for $\phi$, $\theta_\ell$ and $\theta_K$ while the shape is predetermined from the \mbc sideband.
Acceptance and efficiency effects are accounted for in the fit by 
weighting each event by the inverse of its combined efficiency, which is derived from the direct product of the  efficiencies in $\phi$, $\theta_\ell$, $\theta_K$ and  $q^2$.
The individual reconstruction efficiency  for each observable is obtained by extracting the ratio between the reconstructed and generated MC distributions.

All methods  are tested and evaluated in pseudo-experiments using MC samples for each measurement and the results are compared to the input values.
Systematic uncertainties are considered if they introduce an angular- or $q^2$-dependent bias to the distributions of signal or background candidates.
Small correlations between $\theta_\ell$ and $q^2$ are not considered in the treatment of the reconstruction efficiency.
The deviation between a fit  based on generator truth and an MC sample after detector simulation and reconstruction reweighted with efficiency corrections  is evaluated for a bias.
The difference  between the two fits (0.045 on average) is taken as the systematic uncertainty for the efficiency correction; this is the largest systematic uncertainty.
%
Peaking backgrounds are estimated for each $q^2$ bin using MC.
In total, fewer than six (one) such background events  are expected in the muon (electron) channels.
The impact of the peaking component is simulated by performing pseudo-experiments with MC samples for signal and background according to the measured signal yields, replacing six randomly selected events from the signal class with events from simulated peaking background in each measurement. 
The observed deviation from  simulated values ($0.02$ on average) is taken as the systematic uncertainty.
An error on the background parametrization is estimated by repeating all fits with an alternative background description using third-order polynomials and taking the observed deviation (0.028 on average) as the systematic error.
%
Finally, an error on the signal parametrization is considered by repeating the fit with the signal shape parameters adjusted by $\pm 1\sigma$, leading to systematic uncertainties of order $10^{-4}$.
Signal cross-feed is evaluated for all signal decay channels and found to be insignificant.
The parametrization in  Eq.~\ref{eq:signalpdf} does not include a possible S-wave
contribution under the $K^\ast(892)$ mass region. With the expected fraction
of 5\% \cite{lhcb1,lhcb2}, we estimate the S-wave contribution for each measurement
to be less than one event and the resulting effects to be negligible.
Statistically equal numbers of $B$ and $\bar B$ candidates in the signal window are found; consequently,  $\cal CP$-asymmetric contributions to the measured $\cal CP$-even parameters are neglected.
The total systematic uncertainty is calculated as the  sum in quadrature of the individual values.

\begin{figure}
	\centering
	\subfigure{
		\includegraphics[width=\factorh\textwidth]{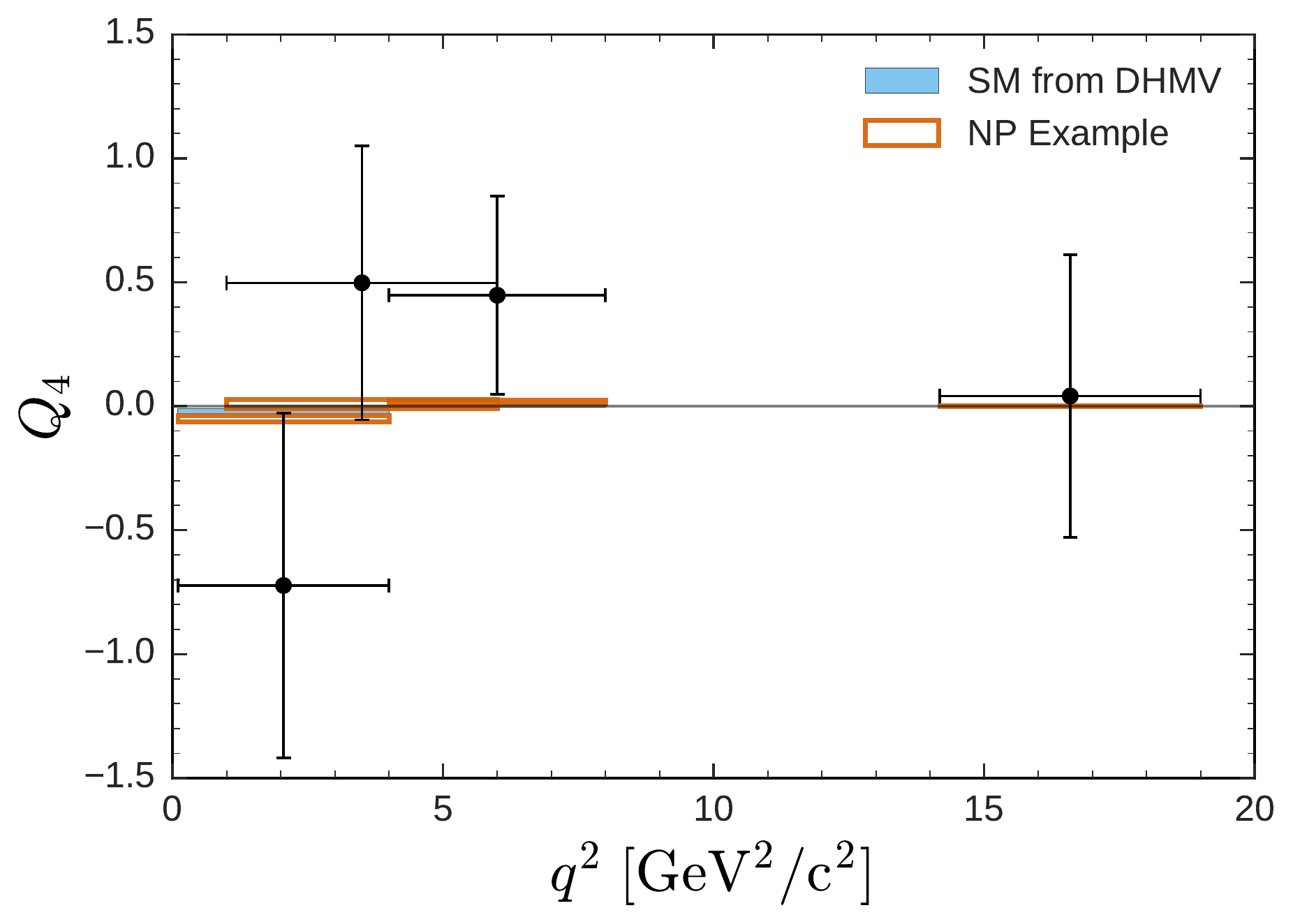}          
	} 
	\subfigure{
		\includegraphics[width=\factorh\textwidth]{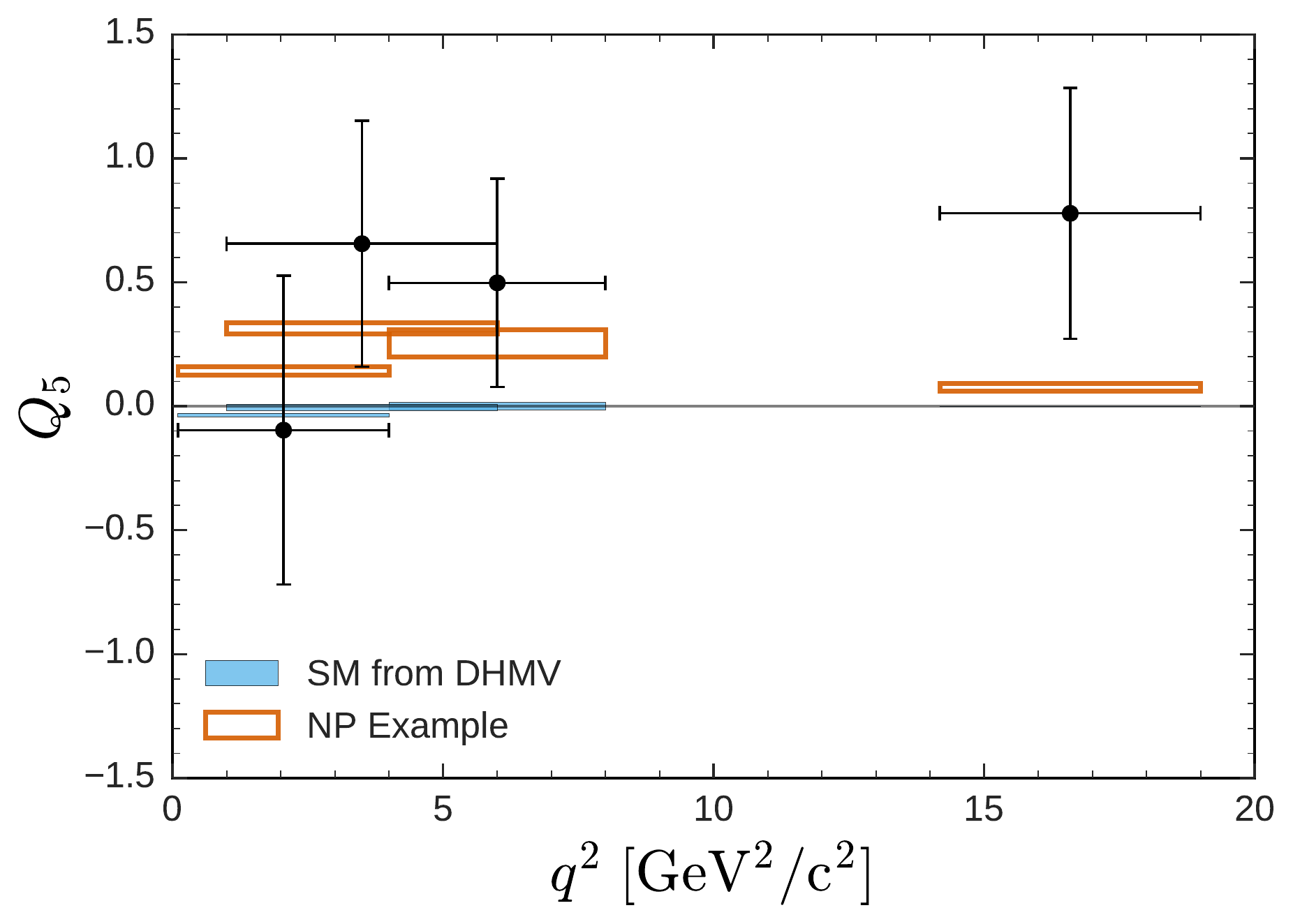}          
	}%
	\caption{$Q_4$ and $Q_5$ observables with SM and favored NP ``Scenario 1" from Ref.~\cite{DHMVeemumu}.}
	\label{fig:resQ}
\end{figure}

\squeezetable
\begin{table*}
	\centering
	\caption{Fit results for $P_4'$ and $P_5'$ for all decay channels and separately for the electron and muon modes. The first uncertainties are statistical 
		and the second  systematic.}
	\label{tab:result}
	\begin{ruledtabular}
	\begin{tabular}{lrrrrrr}
$q^2$ in $\mathrm{GeV}^2/c^2$  &  $P_4'$ &  ${P_4^e}'$&  ${P_4^\mu}'$ &$P_5'$ &  ${P_5^e}'$ &  ${P_5^\mu}'$   \\ [2pt]
\hline
$[1.00 , 6.00]$  &  $-0.45^{+0.23}_{-0.22}\pm 0.09 $   &  $-0.72^{+0.40}_{-0.39}\pm 0.06 $   &  $-0.22^{+0.35}_{-0.34}\pm 0.15 $   &  $0.23^{+0.21}_{-0.22}\pm 0.07 $   &  $-0.22^{+0.39}_{-0.41}\pm 0.03 $   &  $0.43^{+0.26}_{-0.28}\pm 0.10 $  \\ [2pt]
\hline
$[0.10 , 4.00]$  &  $0.11^{+0.32}_{-0.31}\pm 0.05 $   &  $0.34^{+0.41}_{-0.45}\pm 0.11 $   &  $-0.38^{+0.50}_{-0.48}\pm 0.12 $   &  $0.47^{+0.27}_{-0.28}\pm 0.05 $   &  $0.51^{+0.39}_{-0.46}\pm 0.09 $   &  $0.42^{+0.39}_{-0.39}\pm 0.14 $  \\ [2pt]
$[4.00 , 8.00]$  &  $-0.34^{+0.18}_{-0.17}\pm 0.05 $   &  $-0.52^{+0.24}_{-0.22}\pm 0.03 $   &  $-0.07^{+0.32}_{-0.31}\pm 0.07 $   &  $-0.30^{+0.19}_{-0.19}\pm 0.09 $   &  $-0.52^{+0.28}_{-0.26}\pm 0.03 $   &  $-0.03^{+0.31}_{-0.30}\pm 0.09 $  \\ [2pt]
$[10.09 , 12.90]$  &  $-0.18^{+0.28}_{-0.27}\pm 0.06 $   &  -   &  $-0.40^{+0.33}_{-0.29}\pm 0.09 $   &  $-0.17^{+0.25}_{-0.25}\pm 0.01 $   &  -   &  $0.09^{+0.29}_{-0.29}\pm 0.02 $  \\ [2pt]
$[14.18 , 19.00]$  &  $-0.14^{+0.26}_{-0.26}\pm 0.05 $   &  $-0.15^{+0.41}_{-0.40}\pm 0.04 $   &  $-0.10^{+0.39}_{-0.39}\pm 0.07 $   &  $-0.51^{+0.24}_{-0.22}\pm 0.01 $   &  $-0.91^{+0.36}_{-0.30}\pm 0.03 $   &  $-0.13^{+0.39}_{-0.35}\pm 0.06 $  \\ [2pt]
	 \end{tabular}
	\end{ruledtabular}
\end{table*}
\begin{figure}
	\centering
	\subfigure{
		\includegraphics[width=\factorh\textwidth]{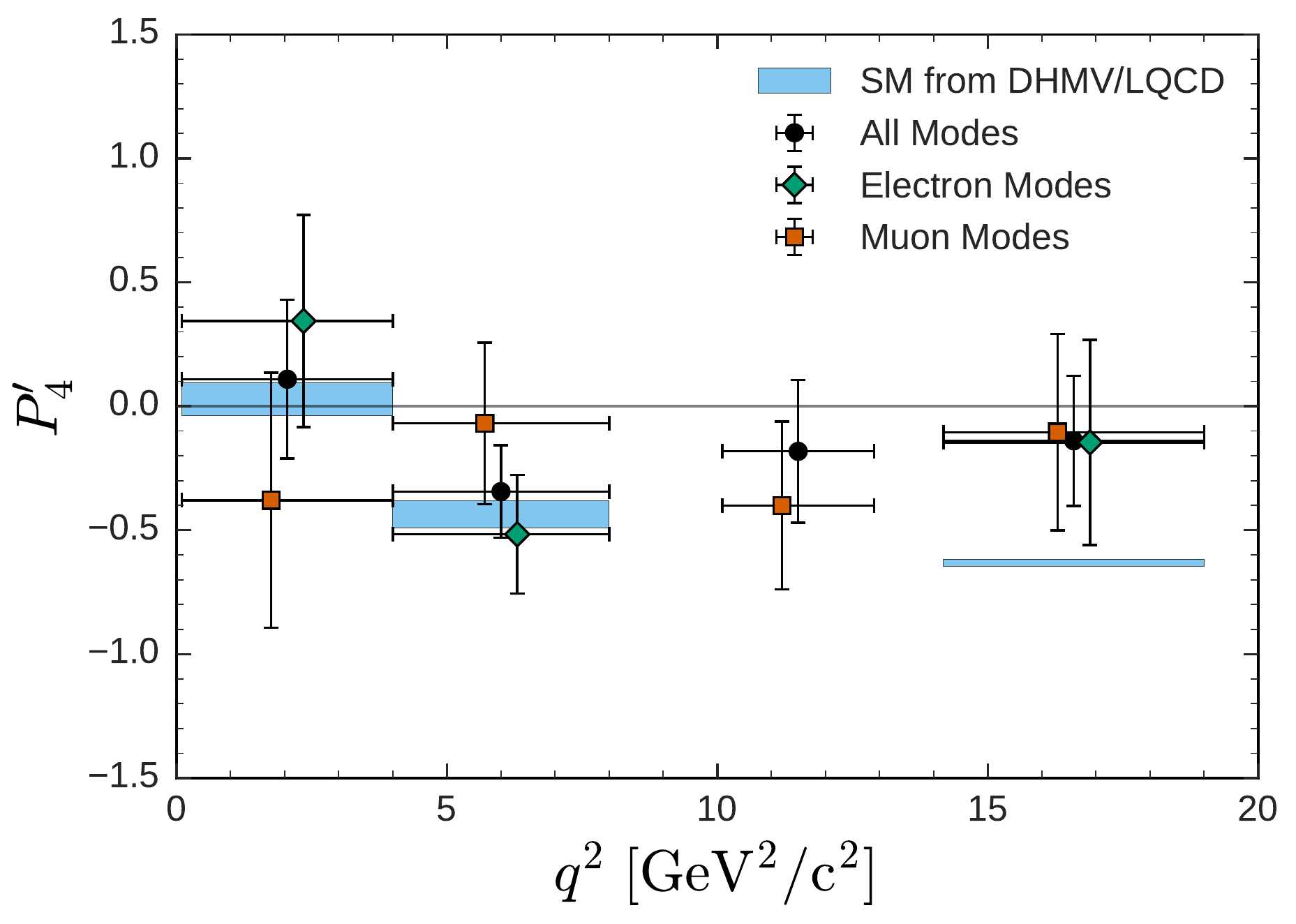}          
	} 
	\subfigure{
		\includegraphics[width=\factorh\textwidth]{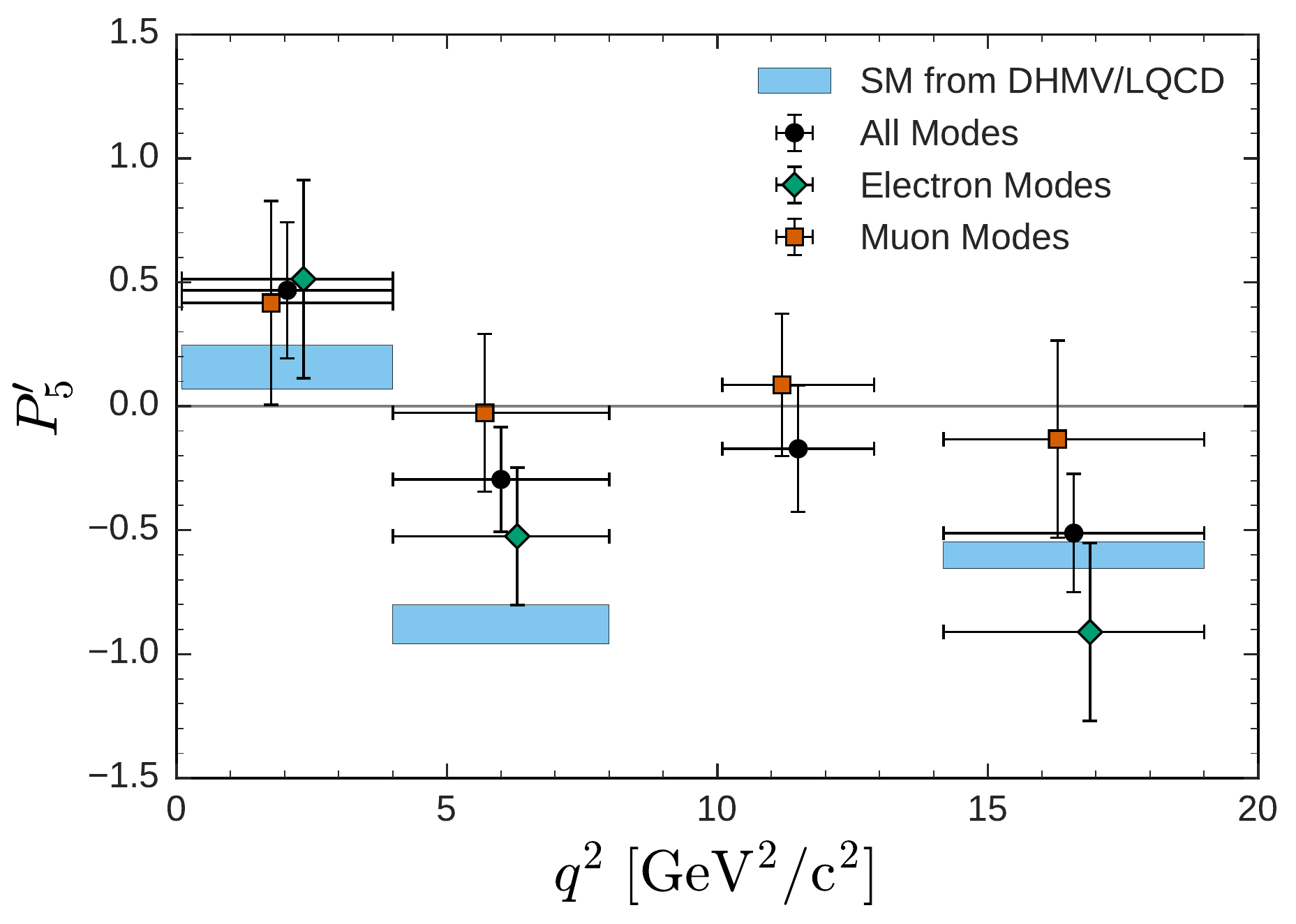}          
	}%
	\caption{$P_4'$ and $P_5'$ observables for  combined, electron and muon modes. 
	The SM predictions are provided by DHMV \cite{DHMVeemumu} and lattice QCD \cite{Horgan:2015vla} and displayed  as boxes   for the muon modes only. The central values of the data points for the  electron and muon modes are shifted horizontally for better readability.}
	\label{fig:res}
\end{figure}
The result of all fits is presented in \Cref{tab:result} and displayed in Fig.~\ref{fig:res} where it is compared to SM predictions by DHMV, 
which  refers  to the soft form-factor method of Ref.~\cite{DHMVmethod}.
Predictions for the $14.18~\mathrm{GeV}^2/c^2< q^2 < 19.00~\mathrm{GeV}^2/c^2$ bin are calculated using lattice QCD with  QCD form factors from Ref.~\cite{Horgan:2015vla}.
%
The predictions include the lepton mass, leading to minor corrections between the SM values for  the electron and muon modes.
For the electron mode, fits in the region  $10.09~\mathrm{GeV}^2/c^2 < q^2 < 12.90~\mathrm{GeV}^2/c^2$ are excluded because it overlaps with the $\psi(2S)$  veto range, leading  to insufficient statistics for stable fit results. 
In total, all measurements are compatible with  SM predictions.
The strongest tension of $2.6\sigma$ (including systematic uncertainty)  is observed in $P_5'$ of the muon modes for the region  $4~\mathrm{GeV}^2/c^2 < q^2 < 8~\mathrm{GeV}^2/c^2$; this is  in the same region where  LHCb  reported the so-called $P_5'$ anomaly \cite{lhcb1,lhcb2}.
In the same region,  the electron modes deviate by $1.3\sigma$ and all channels combined by $2.5\sigma$ (including systematic uncertainty).
All measurements are compatible between lepton flavors.
The $Q_{4,5}$ observables are presented in \Cref{tab:resq} and Fig.~\ref{fig:resQ}, where no significant deviation from zero is discerned.
\squeezetable
\begin{table}
\scriptsize
	\centering
	\caption{Results for the lepton-flavor-universality-violating observables $Q_4$ and $Q_5$. The first uncertainty  is  statistical and the second   systematic.
	}
	\label{tab:resq}
	\begin{ruledtabular}
	\begin{tabular}{lrr}
$q^2$ in $\mathrm{GeV}^2/c^2$  &  $Q_4$ &  $Q_5$ \\ 
\hline
$[1.00 , 6.00]$  &  $0.498\pm{0.527}\pm{0.166} $   &  $0.656\pm{0.485}\pm{0.103} $  \\ 
\hline
$[0.10 , 4.00]$  &  $-0.723\pm{0.676}\pm{0.163} $   &  $-0.097\pm{0.601}\pm{0.164} $  \\ 
$[4.00 , 8.00]$  &  $0.448\pm{0.392}\pm{0.076} $   &  $0.498\pm{0.410}\pm{0.095} $  \\ 
$[14.18 , 19.00]$  &  $0.041\pm{0.565}\pm{0.082} $   &  $0.778\pm{0.502}\pm{0.065} $  \\ 

	 \end{tabular}
	\end{ruledtabular}
\end{table}

In conclusion,  the first lepton-flavor-dependent angular analysis measuring the observables $P_4'$ and $P_5'$  in the  \bkstll decay is reported and the observables $Q_{4,5}$ are shown for the first time.
The results are compatible with  SM predictions, where the largest discrepancy is  $2.6\sigma$  in $P_5'$ for the muon channels.

We thank the KEKB group for excellent operation of the
accelerator; the KEK cryogenics group for efficient
solenoid operations; and the KEK computer group, the
NII, and PNNL/EMSL for valuable computing and
SINET4 network support. We acknowledge support from
MEXT, JSPS, and Nagoya’s TLPRC (Japan); ARC
(Australia); FWF (Austria); NSFC (China); MSMT
(Czechia); CZF, DFG, and VS (Germany); DST (India);
INFN (Italy); MOE, MSIP, NRF, GSDC of KISTI, and
BK21Plus (Korea); MNiSW and NCN (Poland); MES
and RFAAE (Russia); ARRS (Slovenia); IKERBASQUE
and UPV/EHU (Spain); SNSF (Switzerland); NSC and
MOE (Taiwan); and DOE and NSF (USA).

\end{document}